\documentclass[prl,twocolumn,amsmath,amssymb,showpacs]{revtex4}
 
\usepackage{graphicx}
\usepackage{psfrag}
\usepackage{epsfig}
\usepackage{color}
\usepackage{subfigure}
\definecolor{dred}{rgb}{0.7,0.0,0.0}
\usepackage{dcolumn}
\usepackage{bm}

\newcommand{\beginsupplement}{%
        \setcounter{table}{0}
        \renewcommand{\thetable}{S\arabic{table}}%
        \setcounter{figure}{0}
        \renewcommand{\thefigure}{S\arabic{figure}}%
        \setcounter{section}{0}
        \renewcommand{\thesection}{S\arabic{section}}%
        \setcounter{equation}{0}
        \renewcommand{\theequation}{S\arabic{equation}}%
     }

\begin{document}

%
%
\title{ Bicollinear Antiferromagnetic Order, Monoclinic Distortion, \\
and Reversed Resistivity Anisotropy in FeTe as a Result of Spin-Lattice Coupling.}
%
%
%

\author{Christopher B. Bishop}
\author{Adriana Moreo}
\author{Elbio Dagotto}
\affiliation{Department of Physics and Astronomy,University of Tennessee,
Knoxville, TN 37966, USA} 
\affiliation{Materials Science and Technology Division,
Oak Ridge National Laboratory,Oak Ridge, TN 37831, USA}

\date{\today}

\begin{abstract}
{The bicollinear antiferromagnetic order experimentally observed in FeTe
is shown to be stabilized by the coupling 
$\tilde g_{12}$ between monoclinic lattice distortions
and the spin-nematic order parameter with $B_{\rm 2g}$ symmetry, within 
a three-orbital
spin-fermion model studied with Monte Carlo techniques. 
A finite but small value of $\tilde g_{12}$ is required, with a
concomitant lattice distortion compatible with experiments,
and a tetragonal-monoclinic transition strongly first order.
Remarkably, the bicollinear state found here displays a
planar resistivity with the ``reversed'' puzzling anisotropy 
discovered in transport experiments.
Orthorhombic distortions are also incorporated 
and phase diagrams interpolating between
pnictides and chalcogenides are presented.
We conclude that the spin-lattice coupling discussed here 
is sufficient to explain the challenging
properties of FeTe. 
%
} 

\end{abstract}
 

 
\maketitle
 
{\it Introduction.}
The chalcogenide FeTe has long been considered an unusual
member of the iron-based superconductors family~\cite{intro,dainat}.
Angle-resolved photoemission (ARPES) results~\cite{zhang} 
for this material revealed 
substantial mass renormalizations indicative of electrons that are more strongly 
interacting than in pnictides (see also~\cite{akrap}). The absence of
Fermi surface (FS) nesting instabilities was also conclusively 
established~\cite{xia,phlin}.
Moreover, using single-crystal neutron diffraction, the unexpected
presence in FeTe 
of a ``bicollinear'' magnetic state was reported~\cite{bao}.
This exotic antiferromagnetic (AFM) 
state is known as the E-phase in manganites~\cite{CMR}. 
Phenomenological approaches to rationalize the 
bicollinear state rely on Heisenberg
$J_1$-$J_2$-$J_3$ models~\cite{ma} with the constraints 
$J_3>J_2/2$ and $J_2>J_1/2$, implying that the furthest 
distance coupling $J_3$ must be robust. 
Effective spin models~\cite{ma,bernevig} are certainly valid
descriptions after the distortion
occurs, but they do not illuminate on the 
fundamental reasons for the bicollinear 
state stability~\cite{lipscombe,qinlong}.

Upon cooling, experimentally it is known that the bicollinear state is
reached via a robust first-order phase transition~\cite{bao,chen,fobes}, 
with a concomitant tetragonal ($\mathcal{T}_{\rm etra}$) 
to monoclinic ($\mathcal{M}_{\rm ono}$)  lattice distortion.
The reported lattice distortions in Fe$_{\rm 1.076}$Te
and Fe$_{\rm 1.068}$Te are $\delta_M=|a_M-b_M|/(a_M+b_M) \sim 0.007$ ~\cite{bao}
($a_M$ and $b_M$ are
the low temperature lattice parameters in the $\mathcal{M}_{\rm ono}$ notation).
This distortion is comparable to the orthorhombic ($\mathcal{O}_{\rm rth}$) 
lattice distortion in 
BaFe$_2$As$_2$~\cite{huang} $\delta_O=|a_O-b_O|/(a_O+b_O) \sim 0.004$ 
(now with $a_O$ and $b_O$
the low temperature lattice parameters in the $\mathcal{O}_{\rm rth}$ notation). Since
the lattice is considered a ``passenger'' in the properties of the pnictides, it may
be suspected that it also plays a secondary role for chalcogenides.

Contrary to this reasoning, in this publication we argue that the
lattice plays a more fundamental role in FeTe compounds
than previously anticipated. Specifically, we construct a spin-fermion (SF) model where
lattice and spins are coupled in a manner that includes the $\mathcal{M}_{\rm ono}$ distortion
of FeTe. Using Monte Carlo techniques, we found a strong first-order
$\mathcal{T}_{\rm etra}$ to $\mathcal{M}_{\rm ono}$ lattice transition upon cooling, 
as in experiments~\cite{bao}. Moreover, the bicollinear magnetic order
spontaneously arises at the same critical temperature. Furthermore, this is achieved
with a (dimensionless) spin-lattice coupling $\tilde g_{12}\gtrsim 0.10-0.25$ that 
is weak/intermediate in strength. Surprisingly, we also find the same puzzling
{\it reversed} anisotropy in the low temperature resistivity reported recently~\cite{reverse1,reverse2},
with the AFM direction more resistive than the ferromagnetic (FM), 
contrary to the behavior in pnictides.

Our study also includes the spin-lattice coupling $\tilde g_{66}$ 
that favors orthorhombicity although in this case the crystal's geometry
-- with nearest-neighbors (NN) and next-NN (NNN)
hoppings of similar strength and associated FS nesting -- already 
favors the magnetic $(\pi,0)$ 
collinear state even without involving the lattice. Our analysis
allows for an interpolation between pnictides, with collinear order, 
and chalcogenides, with bicollinear order, using the {\it same}
hopping amplitudes, compatible with band structure calculations that
give similar results for both materials~\cite{DFT}.
In fact, we show that the high temperature 
regime displays a FS with the canonical hole
and electron pockets, leading to
the naive assumption that only $\mathcal{O}_{\rm rth}$ 
and $(\pi,0)$ spin order could be stabilized.
However, our calculations explicitly show that 
strong first-order transitions 
can induce a low-temperature state with no precursors at high temperatures.
In other words, in the absence of the
spin-lattice coupling $\tilde g_{12}$ there is {\it no} 
transition to a bicollinear AFM state.





The presence of both itinerant and localized characteristics in neutron 
experiments for Fe$_{1.1}$Te~\cite{igor} suggest that the SF model
provides a proper framework for iron tellurides.
While in our effort the electronic interactions cannot be fully incorporated, the Hund 
coupling, crucial in the SF model, mimics a Hubbard $U$ by reducing double
occupancy at each orbital. The importance of the Hund coupling
has also  been remarked within ARPES~\cite{phlin}. 
In these respects, our study has the
same degree of accuracy as in previous successful descriptions 
of materials such as manganites~\cite{CMR,more-CMR}.


\begin{figure}[thbp]
\begin{center}
\includegraphics[width=6cm,clip,angle=-90]{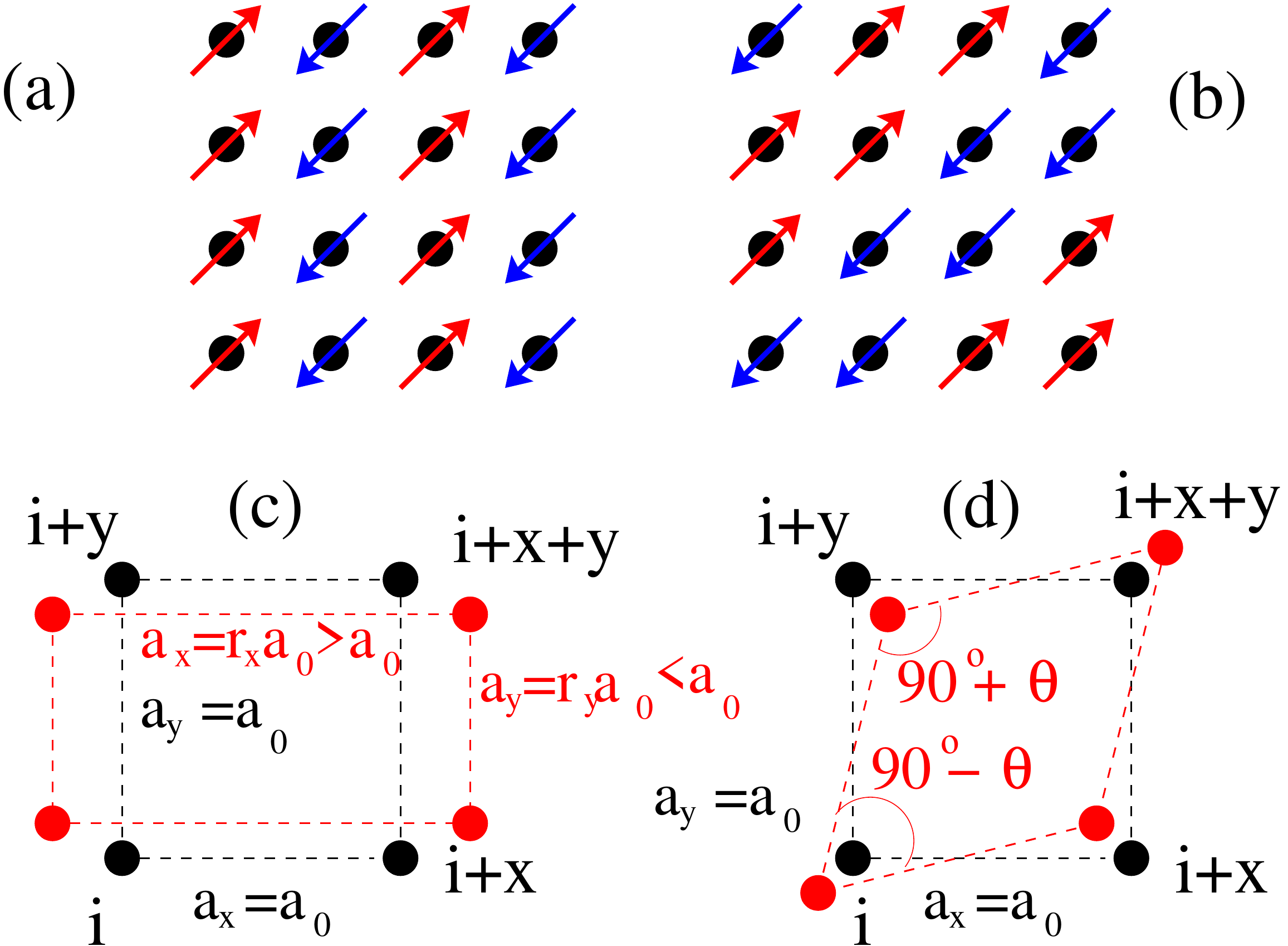}
\vskip -0.3cm
\caption{(color online) (a) The collinear $(\pi,0)$ AFM ordered state;
(b) the bicollinear $(\pi/2,-\pi/2)$ AFM ordered state;
(c) schematic drawing of the Fe lattice equilibrium position
in the $\mathcal{T}_{\rm etra}$ (black symbols) and
$\mathcal{O}_{\rm rth}$ (red symbols) phases 
(four Fe's are indicated with filled circles and labeled by their site index ${\bf i}$); 
(d) Same as (c) but for the $\mathcal{M}_{\rm ono}$ case.
%
}
\label{global}
\end{center}
\end{figure}

{\it Model.} The SF Hamiltonian used here is based
on the original purely electronic model previously discussed~\cite{BNL,shuhua}, 
supplemented by the addition of couplings to the lattice degrees of freedom~\cite{shuhua13,chris}:
\begin{equation}
H_{\rm SF} = H_{\rm Hopp} + H_{\rm Hund} + H_{\rm Heis} + H_{\rm Stiff}+ H_{\rm SLO} + H_{\rm SLM}.
\label{ham}
\end{equation}
\noindent $H_{\rm Hopp}$ is the three-orbital ($d_{xz}$, $d_{yz}$, $d_{xy}$) tight-binding Fe-Fe hopping of electrons, with the
hopping amplitudes selected to reproduce ARPES data [see Eqs.(1-3) and Table 1 
of \cite{three}]. The undoped-limit average electronic density per iron and per orbital 
is $n$=4/3~\cite{three} and 
a chemical potential in $H_{\rm Hopp}$~\cite{chris} controls its value. 
The Hund interaction is $H_{\rm Hund}$=$-{J_{\rm H}}\sum_{{\bf i},\alpha} {{{\bf S}_{\bf i}}\cdot{{\bf s}_{{\bf i},\alpha}}}$,
where ${{\bf S}_{\bf i}}$ are localized spins at site ${\bf i}$  
and ${\bf s}_{{\bf i},\alpha}$ are
itinerant spins corresponding to orbital $\alpha$ at the same site~\cite{foot}.
$H_{\rm Heis}$ contains the NN and NNN 
Heisenberg interactions among the localized spins,
with respective couplings $J_{\rm NN}$
and $J_{\rm NNN}$, and ratio $J_{\rm NNN}$/$J_{\rm NN}$ = 2/3 (any ratio larger than 1/2
leads to similar results below). 
The NN and NNN Heisenberg interactions 
are of comparable magnitude because in FeTe electrons
hop from Fe to Fe via the intermediate Te atom at the center of
Fe plaquettes~\cite{Hubbard}.
$H_{\rm Stiff}$ is the lattice stiffness given by a Lennard-Jones potential 
to speed up convergence~\cite{chris} (see full expression in~\cite{shuhua13}).

Previous SF model investigations focused on the 
$\mathcal{T}_{\rm etra}$-$\mathcal{O}_{\rm rth}$ transition 
as in SrFe$_2$As$_2$~\cite{shuhua13}. 
The coupling of the spins with the $\mathcal{O}_{\rm rth}$ lattice distortion 
discussed in~\cite{shuhua13} is given 
by $H_{\rm SLO}$=$-g_{66}\sum_{\bf i}\Psi^{NN}_{\bf i}\epsilon_{66}({\bf i})$~\cite{fernandes1,fernandes2}, 
where $g_{66}$ is the canonical 
$\mathcal{O}_{\rm rth}$ spin-lattice coupling~\cite{only} and the spin NN
nematic order parameter is defined as 
\begin{equation}
\Psi^{NN}_{\bf i} ={1\over{2}} { {{\bf S}_{\bf i}}\cdot{
({\bf S}_{\bf i+\bf y}+{\bf S}_{\bf i-\bf y}-
 {\bf S}_{\bf i+\bf x}-{\bf S}_{\bf i-\bf x}) }},
\label{PsiNNdef}
\end{equation}
\noindent 
where ${\bf x}$ and ${\bf y}$ are unit vectors along the $x$ and $y$ axes, 
respectively. 
This order parameter is 2 in the perfect ($\pi$,0) state 
shown in Fig.~\ref{global}~(a). 
$\epsilon_{66}({\bf i})$ is the lattice $\mathcal{O}_{\rm rth}$ strain defined in terms of the
positions of the As, Se or Te atoms with respect to their neighboring Fe. Its 
precise definition is~\cite{shuhua13}
\begin{equation}
\epsilon_{66}({\bf i})={1\over{4\sqrt{2}}}\sum_{\nu=1}^4(|\delta^y_{\bf i,\nu}|-|\delta^x_{\bf i,\nu}|),
\label{e66}
\end{equation}
\noindent where $\delta_{\bf i,\nu}=(\delta^x_{\bf i,\nu},\delta^y_{\bf i,\nu})$
($\nu$=1,...,4) is the distance between Fe
at ${\bf i}$ and one of its four neighbors As or Te
(Fig.~S1, Suppl. Sec.~\cite{SuSe}).   
The As/Te atoms are allowed to move locally from their equilibrium position only along
the $x$ and $y$ directions for simplicity. Both
$\Psi^{NN}_{\bf i}$ and $\epsilon_{66}({\bf i})$ transform 
as the $B_{\rm 1g}$ representation of the $D_{\rm 4h}$ group 
under which $H_{\rm SF}$ is invariant.

The crucial novel term 
$H_{\rm SLM}$=$-g_{12}\sum_{\bf i}\Psi^{NNN}_{\bf i}\epsilon_{12}({\bf i})$ introduced
here provides
the coupling between the spin and the $\mathcal{M}_{\rm ono}$ lattice 
distortion~\cite{kuo}. 
The coupling constant is $g_{12}$ and the spin NNN 
nematic order parameter is defined as
\begin{equation}
\Psi^{NNN}_{\bf i} ={1\over{2}} { {{\bf S}_{\bf i}}\cdot{
({\bf S}_{\bf i + \bf x + \bf y}+{\bf S}_{\bf i - \bf x - \bf y}-
 {\bf S}_{\bf i+\bf x - \bf y}-{\bf S}_{\bf i-\bf x+\bf y}) }}.
\label{PsiNNNdef}
\end{equation}
%
\noindent $\Psi^{NNN}_{\bf i}$ becomes 2 in the perfect ($\pi/2$,$-\pi/2$) state shown 
in Fig.~\ref{global}~(b)~\cite{moment}. 
$\epsilon_{12}({\bf i})$ is the lattice $\mathcal{M}_{\rm ono}$ strain 
defined in terms of the Fe-Te/As distances $\delta_{{\bf i},\nu}$ as
\begin{equation}
\epsilon_{12}({\bf i})={1\over{8}}(|\delta_{{\bf i},2}|+|\delta_{{\bf i},4}|-|\delta_{{\bf i},1}|-|\delta_{{\bf i},3}|).
\label{e12}
\end{equation}
$\epsilon_{12}({\bf i})$ transforms as the $B_{\rm 2g}$ representation. For this
reason we must use $\Psi^{NNN}_{\bf i}$, that also transforms as $B_{\rm 2g}$,  
in the product leading to $H_{\rm SLM}$ so that this term is 
invariant under the $D_{\rm 4h}$ group. This simple symmetry argument is the
basic reason for why the bicollinear state is stabilized 
by the monoclinic distortion, as discussed below.

$H_{\rm SF}$ was studied here with the same Monte Carlo (MC) procedure
employed before in~\cite{shuhua13} (details in~\cite{SuSe}). 
The range of couplings for $J_{\rm H}$, $J_{\rm NN}$, and $J_{\rm NNN}$ 
that we used was also extensively discussed before in~\cite{shuhua,shuhua13} 
(see~\cite{SuSe} as well). Our focus instead 
will be on a careful description of the new lattice coupling $\tilde g_{12}$.
During the simulation the As/Te atoms can move locally away from their equilibrium 
positions on the $x$-$y$ plane, while the Fe atoms can move globally in two ways:
(i) via an $\mathcal{O}_{\rm rth}$ distortion characterized by a global 
displacement $(r_x,r_y)$ from the equilibrium position 
$(x^{(0)}_i,y^{(0)}_i)$ of each iron with $r_{\alpha}=1+\Delta_{\alpha}$ ($\Delta_{\alpha}\ll 1$; $\alpha=x$ or $y$) [Fig.~\ref{global}~(c)], and 
(ii) via a $\mathcal{M}_{\rm ono}$ distortion where the angle between two orthogonal Fe-Fe bonds 
is allowed to change globally to $90^o+\theta$ with the four angles
in the $\mathcal{M}_{\rm ono}$ plaquette adding to $360^o$ so that the next 
angle in the plaquette becomes $90^o-\theta$, with $\theta$
a small angle [Fig.~\ref{global}~(d)]. In addition,
the localized (assumed classical) spins ${\bf S_i}$ 
and atomic displacements $(\delta^x_{{\bf i},\nu},\delta^y_{{\bf i},\nu})$ 
that determine the local $\mathcal{O}_{\rm rth}$ or $\mathcal{M}_{\rm ono}$ lattice 
distortion $\epsilon_{66}({\bf i})$~\cite{shuhua13,chris} and $\epsilon_{12}({\bf i})$
are also evaluated via MC. In~\cite{SuSe} we provide the definitions of the spin and lattice 
susceptibilities $\chi_{S(k_x,k_y)}$, $\chi_{\delta_O}$, and $\chi_{\delta_M}$, 
and the dimensionless couplings 
$\tilde g_{66}$ and $\tilde g_{12}$.

\begin{figure}[thbp]
\begin{center}
\includegraphics[width=8cm,clip,angle=0]{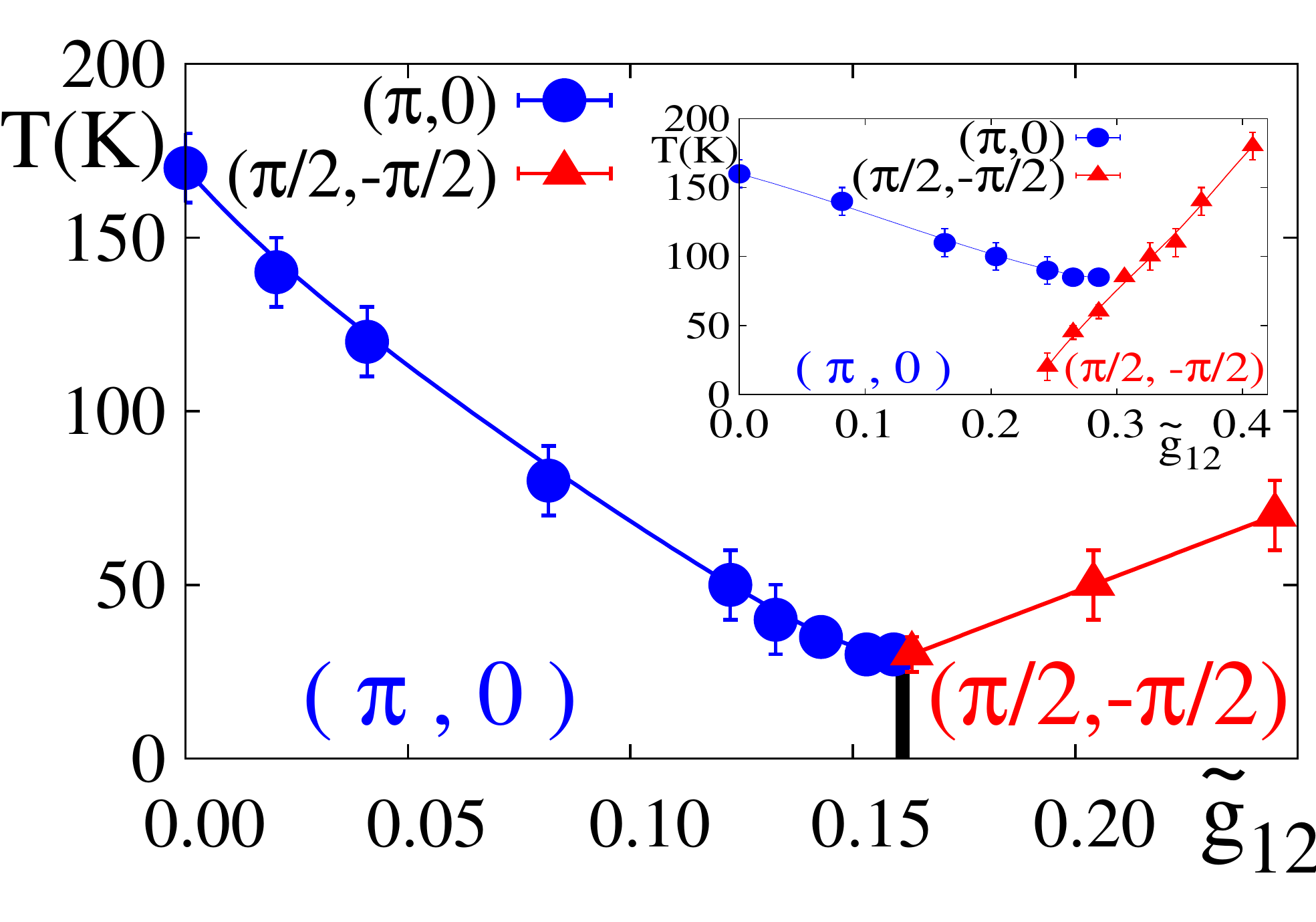}
\vskip -0.3cm
\caption{(color online) Phase diagram varying temperature 
along the straight line from $(\tilde g_{12},\tilde g_{66})=(0,0.24)$ to
$(0.24,0)$, at $J_{\rm H}=0.1$~eV and $J_{\rm NN}$=$J_{\rm NNN}$=0. 
{\it Inset:} the same phase diagram 
but along the straight line from $(\tilde g_{12},\tilde g_{66})=(0,0.16)$ to
$(0.40,0)$, at $J_{\rm H}=0.1$~eV, $J_{\rm NN}=0.012$~eV, and $J_{\rm NNN}=0.008$~eV. 
Blue circles (red triangles) denote $T_{\rm O}$ ($T_{\rm M}$) i.e. 
the transition temperatures to the $\mathcal{O}_{\rm rth}$/collinear 
($\mathcal{M}_{\rm ono}$/bicollinear) phases.}
\label{pdlinevstnh}
\end{center}
\end{figure}

%

{\it Results.}
In real chalcogenides, both $B_{\rm 1g}$ and $B_{\rm 2g}$ magnetic fluctuations 
are expected to be present and the magnitude of their respective 
couplings to $\mathcal{O}_{\rm rth}$ and $\mathcal{M}_{\rm ono}$ distortions may depend on 
doping, replacing Te by Se, or adding extra irons as in Fe$_{1+y}$Te. 
In addition, weak $B_{\rm 2g}$ fluctuations may also exist in 
pnictides. 
%
%
For this reason, our study will be illustrated showing 
the MC phase diagrams varying temperatures and couplings in a wide range. 
Consider first the case $J_{\rm NN}= J_{\rm NNN} = 0.$  
One of our most important results is in Fig.~\ref{pdlinevstnh}.
At the left, a realistic $T_{\rm O}^{max}\approx 170$~K is obtained
for the transition that stabilizes the collinear/$\mathcal{O}_{\rm rth}$ state,
with an $\mathcal{O}_{\rm rth}$ 
distortion $\delta_O\approx 0.004-0.008$, 
compatible with experiments~\cite{bao} 
and with previous studies~\cite{shuhua13}. 
As $~\tilde g_{12}$ increases and $~\tilde g_{66}$ linearly decreases, 
then $T_{\rm O}^{max}$ naturally decreases.
When $~\tilde g_{12} \approx 0.16$ 
and $~\tilde g_{66} \approx 0.08$, remarkably now the FeTe bicollinear/$\mathcal{M}_{\rm ono}$ 
phase appears at  $T_{\rm M}$ (red triangles). 
At the right in Fig.~\ref{pdlinevstnh} 
the critical temperature is $\sim 70$~K
similarly as in FeTe experiments~\cite{mizuguchi}.
Moreover, in the range shown, the monoclinic lattice
distortions are small and compatible with experiments
(for explicit values see Fig.~S4 of~\cite{SuSe})~\cite{comment-10K}.  

Why bicollinear order is stabilized? The reason is that with
increasing $~\tilde g_{12}$, the nematic order parameter
$\Psi^{NNN}_{\bf i}$ in $H_{\rm SLM}$ must develop a nonzero expectation value
to lower the energy. In each odd-even site 
sublattice, $\Psi^{NNN}_{\bf i}$ favors a state with parallel
spins along one diagonal direction and antiparallel in the other 
(equivalent to the collinear order but rotated by 45$^o$). The
parallel locking of the two independent spin sublattices
leads to the state in Fig.~1~(b) (or rotated ones).

As already explained,
the purely fermionic SF model develops a collinear $(\pi,0)$ AFM ground state 
because of FS nesting tendencies in the tight-binding 
sector~\cite{shuhua}. Since spin and lattice are linearly coupled,
an $\mathcal{O}_{\rm rth}$ distortion is induced even for an infinitesimal 
$\tilde g_{66}$.
On the other hand, regardless of $\tilde g_{66}$, we observed that 
the coupling $\tilde g_{12}$ needed 
to stabilize the bicollinear/$\mathcal{M}_{\rm ono}$ state is {\it finite} because it
must first ``fight'' the $(\pi,0)$ order tendencies. However, in practice 
this critical coupling is small $\sim 0.1$-$0.25$ and
within the experimental range.




To analyze the universality of the Fig.~\ref{pdlinevstnh}
phase diagram we have also investigated the effect of adding
NN and NNN Heisenberg couplings along
the line from $(\tilde g_{12},\tilde g_{66})=(0,0.16)$ to $(0.40,0)$
(inset of Fig.~\ref{pdlinevstnh}).
Qualitatively the results are similar.
At $(0.40,0)$ in the inset, the
largest value of $\tilde g_{12}$ considered in the present study, 
the $\mathcal{M}_{\rm ono}$ distortion is $\delta_M\approx 0.004$ still 
of the order of magnitude experimentally observed in FeTe~\cite{bao}. 
One interesting difference, though, between
the two cases in Fig.~\ref{pdlinevstnh} is the appearance in the inset
of an intermediate region at $\tilde g_{12}\approx 0.28$ 
where upon heating a 
transition $\mathcal{M}_{\rm ono}$ to $\mathcal{O}_{\rm rth}$ is reached before the system 
eventually becomes paramagnetic. 
Experimentally it is indeed known that 
in Fe$_{1+y}$Te an intermediate $\mathcal{O}_{\rm rth}$ phase with incommensurate 
magnetic order exists between the $\mathcal{T}_{\rm etra}$ and $\mathcal{M}_{\rm ono}$ 
phases~\cite{mizuguchi,rodriguez} with $T_{\rm O}~\approx 60$~K 
and $T_{\rm M}~\approx 50$~K, at $y\approx 0.13$. 
Although our finite lattices do not have enough resolution 
to study the subtle incommensurate magnetism in detail, we conjecture
that the addition of Fe to FeTe may effectively increase 
the spin-lattice constant values to reach the intermediate regime in the inset.


\begin{figure}[thbp]
\begin{center}
\includegraphics[width=8cm,clip,angle=0]{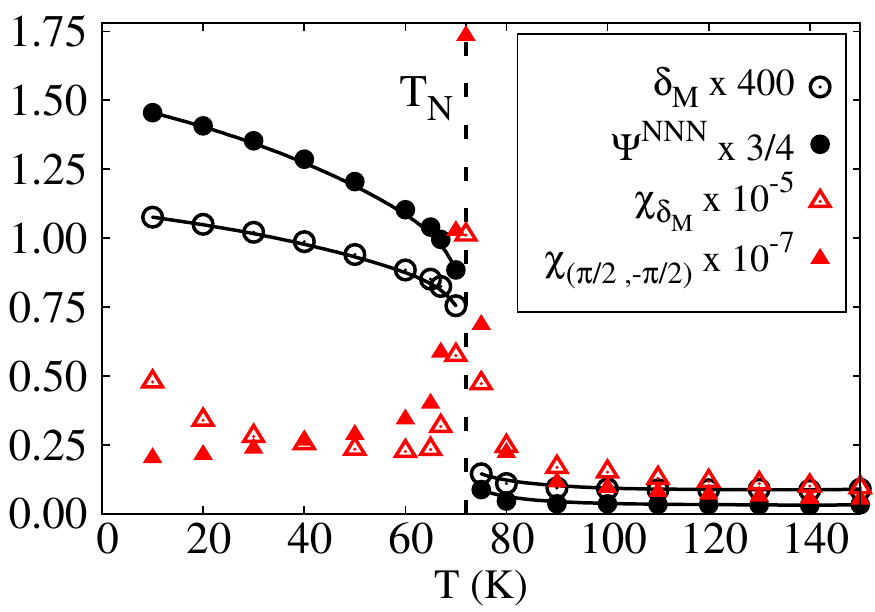}
\vskip -0.3cm
\caption{(color online) 
Filled (open) circles indicate the bicollinear AFM order parameter $\Psi^{NNN}$
(the $\mathcal{M}_{\rm ono}$ lattice distortion $\delta_M$) at 
$\tilde g_{12} = 0.24$, $\tilde g_{66} = 0$, $J_{\rm H}=0.1$~eV, 
and $J_{\rm NN}=J_{\rm NNN}=0$. Magnetic and lattice susceptibilities, 
$\chi_{(\pi/2,-\pi/2)}$ and $\chi_{\delta_M}$, are also shown 
(filled and open triangles, respectively).
$T_{\rm N}$ denotes the first-order Ne\'el temperature.
}
\vskip -0.4cm
\label{sus-noNHg120.24g660}
\end{center}
\end{figure}

\begin{figure}[thbp]
\begin{center}
\includegraphics[width=8cm,clip,angle=0]{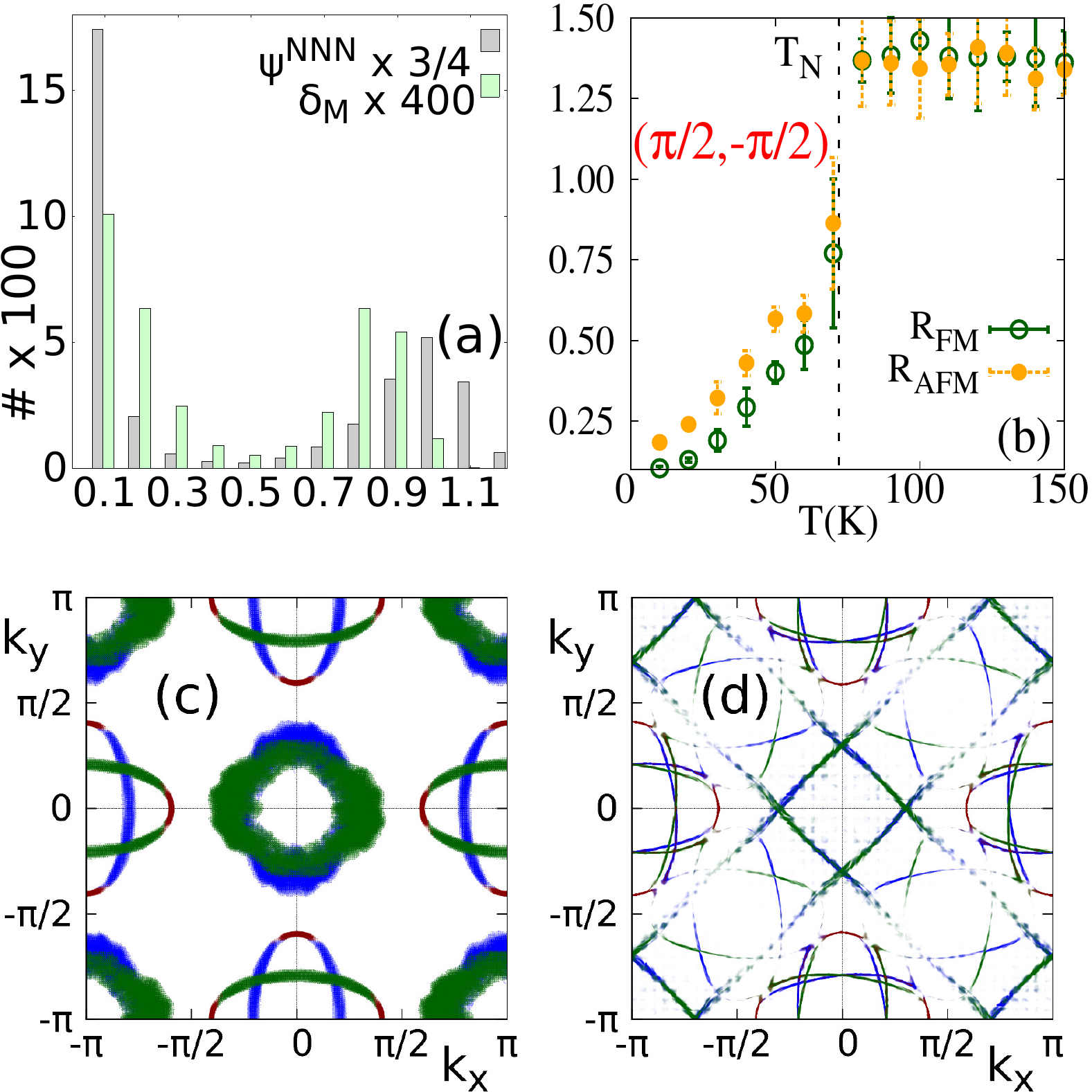}
\vskip -0.3cm
\caption{(color online) 
(a) Histogram of the MC time evolution of $\Psi^{NNN}$ and $\delta_M$, 
at the critical temperature of Fig.~\ref{sus-noNHg120.24g660} ($T=72$~K), 
illustrating its bimodal character compatible with first-order characteristics.
(b) Resistance ($h/2e^2$ units) vs. temperature in the bicollinear state 
($\tilde g_{12}=0.24$, $\tilde g_{66}=0$, $J_{\rm H}=0.2$~eV, no Heisenberg terms). 
Filled (open) symbols denote resistivities along the AFM (FM) direction.
(c,d) Symmetrized Fermi surface 
($\tilde g_{12}=0.24$, $\tilde g_{66}=0$, $J_{\rm H}=0.2$~eV, no Heisenberg terms).  
(c) is in the high temperature paramagnetic phase 
($T=360$~K); (d) is in the bicollinear phase ($T=10$~K). The FS orbital 
composition notation is blue ($xz$), green ($yz$), and red ($xy$). In the
non-symmetrized FS (not shown) a gap opens along the AFM diagonal direction in 
the $xz$ and $yz$ orbitals, compatible with the resistivity results.
}
\vskip -0.4cm
\label{RandFS}
\end{center}
\end{figure}

Another  important result unveiled here is that 
the bicollinear/$\mathcal{M}_{\rm ono}$ phase transition was found to be 
strongly first order, in agreement with experiments~\cite{bao},
as indicated by the order parameters 
discontinuities shown in Fig.~\ref{sus-noNHg120.24g660} and by
the MC-time evolution histogram shown in Fig.~\ref{RandFS}~(a). 
The reason is that at high temperature $(\pi,0)$ fluctuations first 
develop (as implied by the inset of Fig.~2), 
leading to a free energy local minimum. 
However, upon further cooling the bicollinear minimum with a different symmetry
also develops and eventually a crossing occurs with first-order characteristics
because one local state cannot evolve smoothly into the other.




Remarkably, in addition to reproducing properly the FeTe 
structural/magnetic transitions, the correct behavior 
for the resistivity anisotropy~\cite{reverse1,reverse2} is also 
observed. In the $(\pi,0)$ collinear phase, 
FS nesting opens a pseudogap for the $yz$ orbital~\cite{shuhua13,shuhua,PG-our}. 
Because this orbital relates to electronic hopping along the ferromagnetic $y$-axis,
then the FM resistivity is the largest in pnictides. However, the
{\it reversed} anisotropy with lower resistance along 
the FM direction (open circles) was found 
in the bicollinear phase Fig.~\ref{RandFS}~(b)
(the technique used was explained in~\cite{shuhua}).
Moreover, we have 
noticed that this reversed effect is amplified 
as $J_{\rm H}$ increases. The key clues to explain the
effect are now clear: (i) when an electron hops along the plaquette diagonal 
in the AFM direction  it pays
an energy $J_{\rm H}$, but the hopping along the plaquette diagonal 
FM direction does not
have this penalization; (ii) because FS nesting does not involve 
wavevectors such as $(\pi/2,-\pi/2)$, then pseudogaps are not created 
due to nesting as in pnictides. Then, in essence, the reversed resistance
found here is characteristic of physics 
of large Hund coupling materials~\cite{hund-aniso}, such
as manganites~\cite{CMR}, where it is also known that the AFM direction 
is more resistive than the FM direction.

A paradox of FeTe is that first principles studies predict
FS nesting and, thus, $(\pi,0)$ order as in pnictides. For this reason,
we calculated the FS at couplings where the ground state is $\mathcal{M}_{\rm ono}$. 
Figure \ref{RandFS}~(c) shows 
the FS in the high temperature $\mathcal{T}_{\rm etra}$ state.
It is similar to that of the iron pnictides, thus suggestive of
$(\pi,0)$ order upon cooling (the $\Gamma$ centered 
features are blurry because of how a shallow pocket 
is affected by temperature). However, as shown before, because of the sharp 
first-order transition
the $\mathcal{M}_{\rm ono}$ state reached at low temperature has a peculiar FS
 [Fig.~\ref{RandFS}~(d)]: while the electron pockets are similar, 
the squarish $\Gamma$ hole pocket is different
from that of pnictides. In addition ``shadow bands'' features at 
$(\pm \pi/2,\pm \pi/2)$ develop, as observed in ARPES~\cite{xia}, indicative of
couplings stronger than for pnictides. 

{\it Discussion.} Using computational techniques applied to the
spin-fermion model including a spin-lattice
$\mathcal{M}_{\rm ono}$ distortion in the $B_{\rm 2g}$ channel, 
we show that the (often puzzling) 
phenomenology of FeTe can be well reproduced. 
This includes the presence of bicollinear magnetic
order and $\mathcal{M}_{\rm ono}$ lattice distortions,
a strong first-order $\mathcal{T}_{\rm etra}$-$\mathcal{M}_{\rm ono}$ transition,
Fermi surfaces at high temperature that naively would favor 
$(\pi,0)$ magnetic order, and last but not least also
the low-temperature reversed anisotropic resistances between the AFM and FM 
directions. Moreover, all this is achieved with
spin-lattice dimensionless couplings substantially less than 1, 
and with associated lattice distortions 
$\delta_M \sim 10^{-3}$ as in FeTe
experiments. 

While in pnictides the resistance anisotropy is related to FS nesting and
a pseudogap in the $yz$ orbital~\cite{PG-our}, here we argue that in chalcogenides 
the strength of the Hund coupling is more
important for transport since the reversed anisotropy increases with $J_{\rm H}$.
To our knowledge, the spin-lattice interaction discussed here provides
the first physical explanation of a vast array of experimental challenging 
results in FeTe.




{\it Acknowledgments.}
Discussions with S. Liang, P. Dai, and J. Tranquada are acknowledged.
C.B. was supported by the National Science Foundation, under
Grant No. DMR-1404375. E.D. and A.M. were supported by the US Department of Energy, 
Office of Basic Energy Sciences, Materials Sciences and Engineering
Division.


\beginsupplement

\vfill

\begin{center}
\textbf{\LARGE SUPPLEMENTAL\\
\vspace{1 mm} 
MATERIAL}\\
\vspace{1 mm} 
\end{center}


In this supplemental section, technical details and additional results are provided.
 
\section {LATTICE DISPLACEMENTS}

\begin{figure}[thbp]
\begin{center}
\includegraphics[width=4cm,clip,angle=-90]{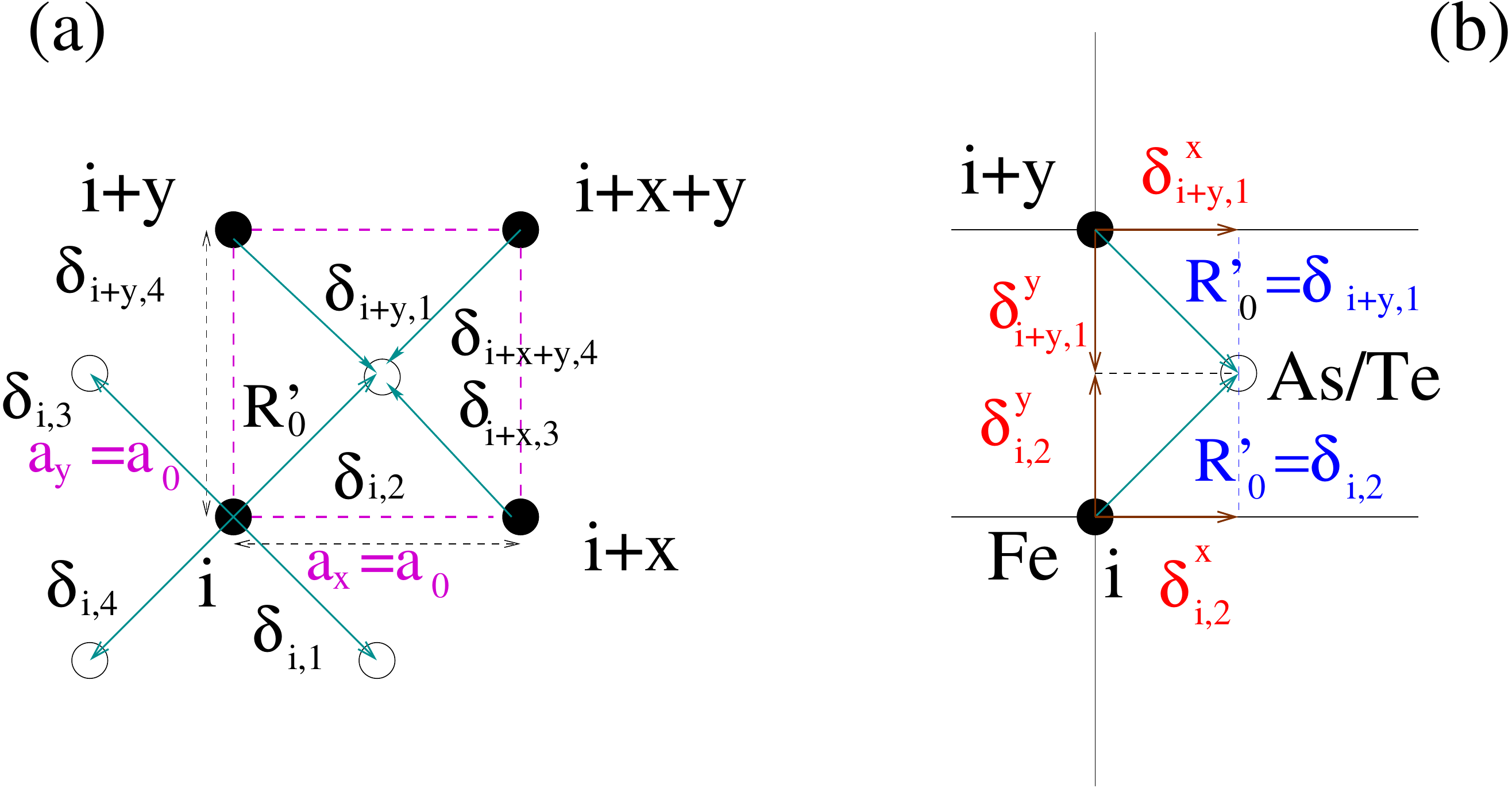}
\vskip 0.3cm
\caption{(color online) (a) Schematic representation 
of the equilibrium position of the Fe-Te/As  lattice 
(projected on the $x$-$y$ plane). Four Fe atoms are 
indicated with filled circles and labeled by their site 
index ${\bf i}$. The open circles indicate the projection of the 
equilibrium position of the As/Te atoms on the $x$-$y$ plane. The 
distance between an Fe atom at site ${\bf i}$ and its 
four neighboring As/Te atoms is indicated by 
$\delta_{{\bf i},\nu}$ with $\nu$ 
running from 1 to 4 (turquoise arrows). 
In equilibrium $\delta_{{\bf i},\nu}=\sqrt{2}a_0/2$. 
The dashed lines indicate $a_x=a_y=a_0$, the equilibrium distance between neighboring irons. 
(b) Sketch representing the variables $\delta_{{\bf i},\nu}^x$ and $\delta_{{\bf i},\nu}^y$ 
(brown arrows) for labels (${\bf i},2$) and (${\bf i+y},1$) 
in the equilibrium configuration. 
For an illustration of the non-equilibrium 
$\delta_{{\bf i},\nu}$ see~\cite{shuhua13}.}
\label{lattice}
\end{center}
\end{figure}

The lattice variables $\delta_{\bf i,\nu}=(\delta^x_{\bf i,\nu},\delta^y_{\bf i,\nu})$,
with $\nu$ ranging from 1 to 4, that enter in the definition 
of $\epsilon_{66}$ and $\epsilon_{12}$, the orthorhombic and monoclinic lattice distortions
respectively, represent the distance between
an Fe atom at site ${\bf i}$ (filled circles in Fig.~\ref{lattice}) 
and one of its four neighboring As or Te atoms 
(open circles in the figure and labeled by the index $\nu$).   
The As/Te atoms are allowed to move locally from their equilibrium position, 
but only along the directions $x$ and $y$ (the $z$ 
coordinate does not participate in the planar lattice distortions addressed here).

\section {METHODS}

The Hamiltonian $H_{\rm SF}$ defined in the main text was studied using a 
Monte Carlo method~\cite{shuhua,CMR} applied to 
{\it (i)} the localized spin 
degrees of freedom ${\bf S_i}$ assumed classical, {\it (ii)} the atomic displacements 
$(\delta^x_{{\bf i},\nu},\delta^y_{{\bf i},\nu})$ that determine the local 
orthorhombic or monoclinic lattice distortions 
$\epsilon_{66}({\bf i})$~\cite{shuhua13,chris} and $\epsilon_{12}({\bf i})$, 
{\it (iii)} the global orthorhombic distortion $(r_x,r_y)$, and {\it (iv)} the 
global monoclinic distortion $\theta$.
As already explained, in the MC simulation the As/Te atoms are 
allowed to move from their equilibrium positions on the $x-y$ plane 
but the Fe atoms can only move globally in two ways:
(i) via a global orthorhombic distortion characterized 
by a global displacement $(r_x,r_y)$ from the equilibrium position 
$(x^{(0)}_i,y^{(0)}_i)$ of each Fe atom, with $r_{\alpha}=1+\Delta_{\alpha}$ 
($\Delta_{\alpha}\ll 1$) and $\alpha=x$ or $y$ [see panel
(c) of Fig.~1, main text];
(ii) via the angle between two orthogonal Fe-Fe bonds which is 
allowed to change globally to $90^o+\theta$ with the four angles
in the monoclinic plaquette adding to $360^o$ so that the following 
angle in the plaquette becomes $90^o-\theta$, with $\theta$
a small angle [see panel (d) of Fig.~1, main text].  After the global 
distortion the new position of the Fe atom is given by
\begin{equation}
\biggl\{
\begin{array}{ll}
 x_i=x^{(0)}_ir_x\cos\theta-y^{(0)}_ir_y\sin\theta \\
 y_i=-x^{(0)}_ir_x\sin\theta+y^{(0)}_ir_y\cos\theta. 
\end{array}
\label{r}
\end{equation}
When an orthorhombic distortion is stabilized, the variables $\delta^s_{\bf i,\nu}$ 
satisfy the constrain
\begin{equation}
2Na_s=\sum_{{\bf i}=1}^N\sum_{\nu=1}^4|\delta^s_{\bf i,\nu}|,
\label{deltao}
\end{equation}
\noindent where $N$ is the number of Fe sites, 
$s=x,y$, and $a_s=a_0r_s$ is the constant Fe-Fe distance along the $s$ direction 
which is equal to $a_0$ in the undistorted tetragonal phase as shown
in panel (c) of Fig.~1 (main text). 
The orthorhombic distortion order parameter $\delta_O$ is then given by
\begin{equation}
\delta_O={|a_x-a_y|\over{a_x+a_y}}={a_0|r_x-r_y|\over{a_0(r_x+r_y)}}.
\label{deltaO}
\end{equation}

Since $r_s=1+\Delta_s$ and $s=x$, $y$, then
\begin{equation}
\delta_O={|1+\Delta_x-(1+\Delta_y)|\over{1+\Delta_x+1+\Delta_y}}\approx{1\over{2}}{|\Delta_x-\Delta_y|}.
\label{deltaOD}
\end{equation}
  
On the other hand, when a monoclinic distortion is stabilized 
the constraint satisfied by $\delta^s_{\bf i,\nu}$ is given by 
\begin{equation}
2Nd_{x+y}=\sum_{{\bf i}=1}^N(|\delta_{{\bf i},4}|+|\delta_{{\bf i},2}|),
\label{deltam1}
\end{equation}
\noindent and
\begin{equation}
2Nd_{x-y}=\sum_{{\bf i}=1}^N(|\delta_{{\bf i},3}|+|\delta_{{\bf i},1}|),
\label{deltam2}
\end{equation}
\noindent where $d_{\mu}$ is the length of the plaquette's diagonal along 
the $\mu$ direction of the plaquette formed by four Fe atoms. In the tetragonal 
phase $d_{\mu}=\sqrt{2}a_0$ while in the monoclinic phase 
$d_{\mu}=\sqrt{2}a_0\sqrt{1-\cos(90^o\pm\theta)}$ with the 
plus (minus) sign for $\mu=x-y$ ($x+y$) 
[see panel (d) of Fig.~1, main text]. The monoclinic 
distortion order parameter $\delta_M$ is then given by
\begin{multline}
\delta_M={|d_{x+y}-d_{x-y}|\over{d_{x+y}+d_{x-y}}}=\\
{\sqrt{2}a_0|(1-\sin\theta)^{1/2}-(1+\sin\theta)^{1/2}|\over{\sqrt{2}a_0((1-\sin\theta)^{1/2}+(1+\sin\theta)^{1/2})}}\approx
{\theta\over{2}}.
\label{deltaM}
\end{multline}
In summary, Monte Carlo simulations are performed on the 
values for the lattice variables $r_x$, $r_y$, $\theta$, and $\delta^s_{{\bf i},\nu}$, 
and also on the localized spin variables ${\bf S_i}$.
 
For each fixed 
Monte Carlo configuration of spins, atomic positions and global distortions, the 
remaining quantum fermionic Hamiltonian is diagonalized. The simulations were performed 
varying the temperature $T$ and the spin-lattice dimensionless couplings
$\tilde g_{66}$  and $\tilde g_{12}$. The latter 
are defined by $\tilde g_{66}={2g_{66}\over{\sqrt{kW}}}$ 
and $\tilde g_{12}={2g_{12}\over{\sqrt{kW}}}$ where $W=3$~eV is the
bandwidth of the tight-binding portion of the Hamiltonian 
and $k$ is a constant that appears in $H_{\rm Stiff}$ (for details see~\cite{shuhua13}). 
The range of values explored for these dimensionless coupling constants 
was chosen so that the orthorhombic and monoclinic distortions (also dimensionless defined) 
agree with the experimental values that range 
from 0.003 to 0.007~\cite{bao,li,huang}.

The fermionic exact diagonalization technique results can be obtained 
comfortably only on up to $8\times 8$ lattices which is the cluster 
size used in this work. However, 
twisted boundary conditions were also used~\cite{salafranca} in the evaluation 
of the resistivities and Fermi surfaces (FS),  
effectively increasing the lattice size as explained in early efforts~\cite{shuhua13}. 
Most couplings were fixed to values used
successfully in previous investigations~\cite{shuhua} for simplicity: 
${J_{\rm H}}$=$0.1$~eV, ${J_{\rm NN}}$=$0.012$~eV, and ${J_{\rm NNN}}$=$0.008$~eV. However, results for 
${J_{\rm H}}$=$0.2$~eV and ${J_{\rm NN}}$=${J_{\rm NNN}}$=0 were 
also discussed in the main text. 

In the Monte Carlo simulations typically 5,000 MC lattice sweeps were used for  
thermalization and 10,000 to 25,000 for 
measurements, at each temperature and parameter values investigated.  
In addition to the $B_{\rm 2g}$ order parameter, 
the magnetic transition was also determined from the behavior of 
the magnetic susceptibility defined as
\begin{equation}
\chi_{S(k_x,k_y)}=N\beta\langle S(k_x,k_y)-\langle S(k_x,k_y)\rangle\rangle^2,
\label{Xs}
\end{equation}
\noindent where $\beta=1/k_BT$, $N$ is the number of lattice sites, 
and $S(k_x,k_y)$ is the magnetic structure factor at wavevector $(k_x,k_y)$ 
obtained via the Fourier transform of the real-space spin-spin 
correlations measured in the MC simulations. To study the 
collinear [bicollinear] AFM state $(k_x,k_y)$ was set to $(\pi,0)$ [$(\pi/2,-\pi/2)$].

Besides the lattice order parameter $\delta_O$ 
given in Eq.~\ref{deltaO}, 
the orthorhombic structural transition was determined from the behavior 
of the lattice susceptibility defined as
\begin{equation}
\chi_{\delta_O}=N\beta\langle \delta_O-\langle\delta_O\rangle\rangle^2.
\label{Xa}
\end{equation}

Reciprocally, the monoclinicic structural transition was studied via
its order parameter, i.e. the monoclinic distortion $\delta_M$ given in 
Eq.~\ref{deltaM}, and also through 
the lattice susceptibility defined as
\begin{equation}
\chi_{\delta_M}=N\beta\langle \delta_M-\langle\delta_M\rangle\rangle^2.
\label{Xthe}
\end{equation}

\section {ADDITIONAL PHASE DIAGRAMS}

\begin{figure}[thbp]
\begin{center}
\includegraphics[trim = 0mm 0mm 0mm 0mm,width=\linewidth,clip,angle=0]{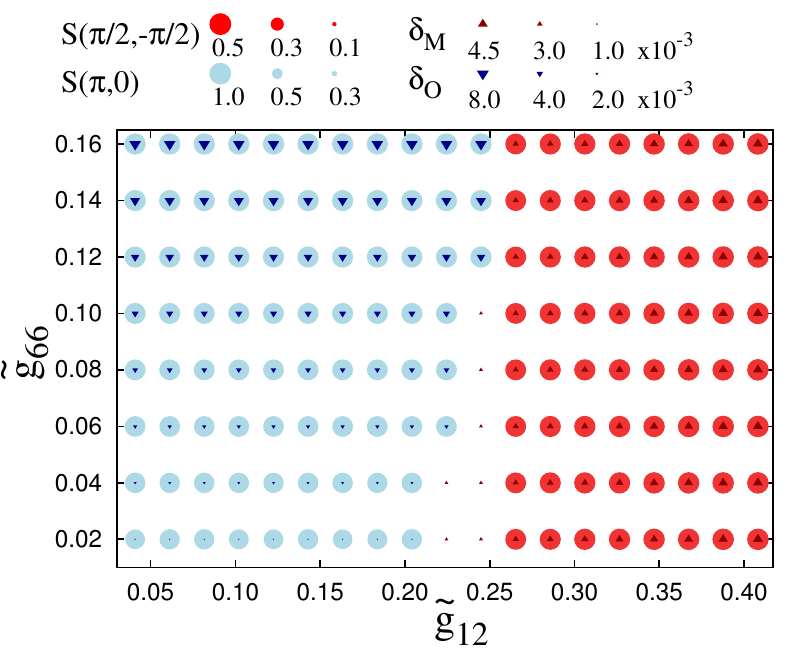}
\vskip -0.3cm
\caption{(color online) Phase diagram at $T=10$~K of the spin fermion 
model, including Heisenberg couplings with the values indicated in the Methods Section,  
varying the dimensionless couplings to the orthorhombic and monoclinic 
distortions. The size of the blue (red) circles is 
proportional to the strength of the collinear (bicollinear) AFM order. 
The size of the bottom side up (down) triangles is proportional 
to the magnitude of the orthorhombic (monoclinic) distortion. The actual scales
used are shown at the top of the figure.
}
\label{pdt10}
\end{center}
\end{figure}

The phase diagram as a function of the 
couplings $\tilde g_{66}$  and $\tilde g_{12}$ at $T=10$~K 
is presented in Fig.~\ref{pdt10} including Heisenberg couplings. It is important 
to remember that in the absence of spin-lattice couplings the SF model already develops 
a collinear AFM ground state due to the comparable
NN and NNN hoppings in the tight-binding term of the Hamiltonian 
(and the concomitant NN and NNN Heisenberg interactions 
between the localized spins if included~\cite{shuhua}). The coupling
$\tilde g_{66}$ that couples the short-range $B_{\rm 1g}$ magnetic nematic 
operator to the orthorhombic distortion stabilizes a small orthorhombic 
distortion that increases monotonically 
with the value of this spin-lattice coupling, as indicated by the size of the 
inverted triangles in the figure. The blue circles indicate 
the concomitant presence of collinear $(\pi,0)$ AFM order. 
The figure shows that, regardless of $\tilde g_{66}$, 
the coupling $\tilde g_{12}$, between the monoclinic lattice distortion 
and the $B_{\rm 2g}$ magnetic nematic operator, 
has to reach a finite value close to 0.25 
to stabilize the bicollinear AFM state indicated by the red circles in the figure. 
The bicollinear magnetic order is accompanied by a monoclinic 
lattice distortion indicated by the triangles 
whose size increases monotonically with $\tilde g_{12}$.

It is interesting to observe that there is a region in the phase diagram 
Fig.~\ref{pdt10} where the monoclinic distortion is stabilized, but the magnetic order 
is neither collinear nor bicollinear. This is caused by
the competition between $\tilde g_{12}$, that after inducing 
the monoclinic distortion induces the bicollinear magnetic order, 
and the NN and NNN Heisenberg couplings that favor a collinear $(\pi,0)$
magnetic state. Thus, $\tilde g_{12}$ is able to induce the lattice 
distortion before it clearly stabilizes the bicollinear magnetic order. 
The fact that the value of $\tilde g_{12}$ that stabilizes the 
bicollinear state is larger than the value of $\tilde g_{66}$ needed 
to obtain the experimentally observed magnitude of the orthorhombic 
distortion is also a result of the effect of the Heisenberg terms in the 
Hamiltonian that favor the collinear AFM state. 

\begin{figure}[thbp]
\begin{center}
\includegraphics[trim = 0mm 0mm 0mm 0mm,width=\linewidth,clip,angle=0]{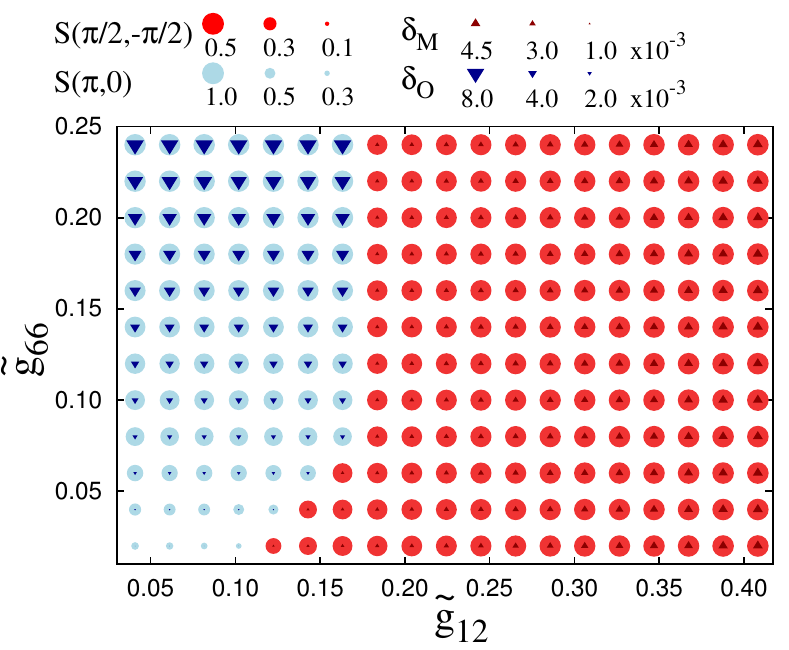}
\vskip -0.3cm
\caption{(color online) Phase diagram at $T=10$~K 
corresponding to the spin-fermion model for the case $J_{\rm NNN}$=$J_{\rm NN}$=0,  
varying the spin-lattice couplings  that lead to the orthorhombic 
and monoclinic distortions. The size of the blue (red) circles is 
proportional to the strength of the collinear (bicollinear) AFM order, 
while the size of the bottom side up (down) triangles is proportional 
to the magnitude of the orthorhombic (monoclinic) distortion.}
\label{pdt110NH}
\end{center}
\end{figure}

\begin{figure}[thbp]
\begin{center}
\includegraphics[width=8cm,clip,angle=0]{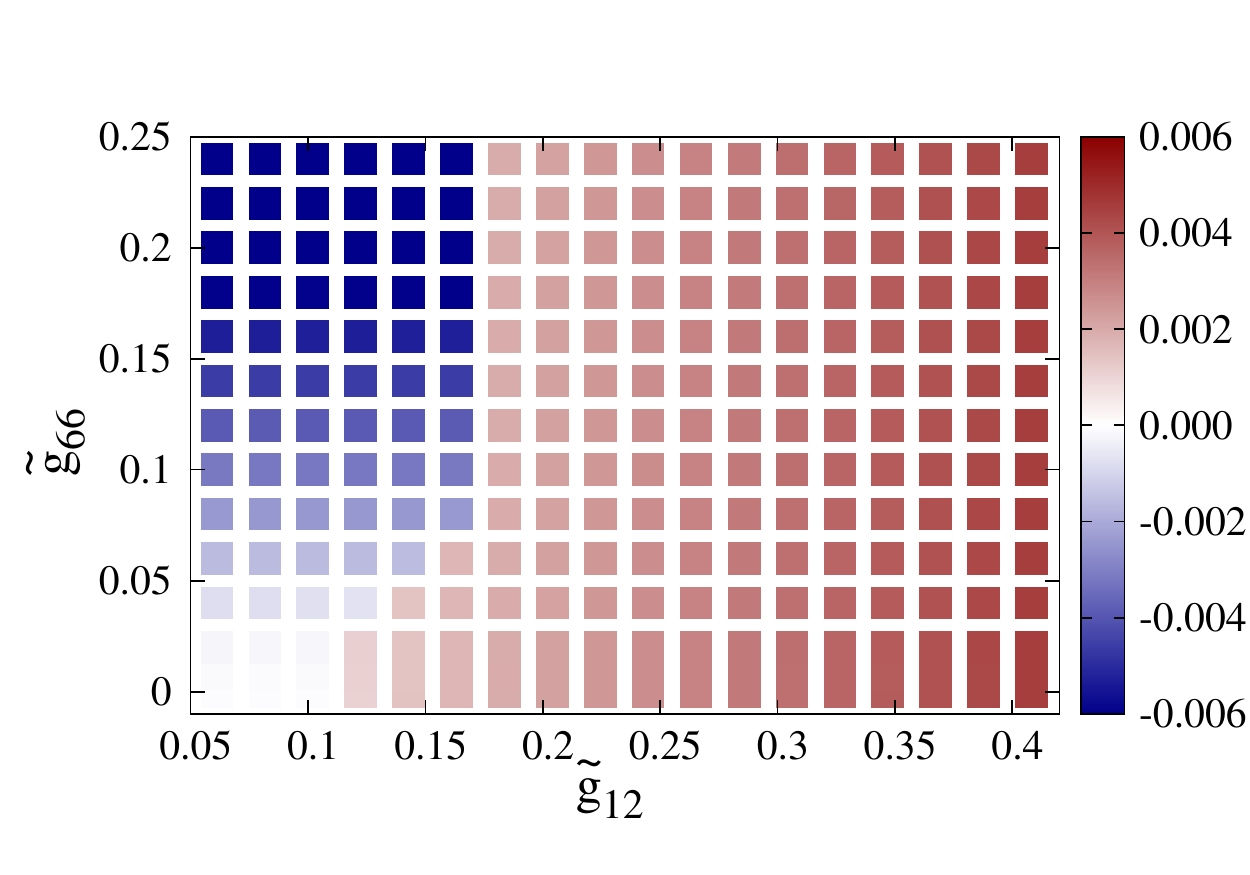}
\vskip -0.3cm
\caption{(color online) Orthorhombic, $\delta_O$ (blue), and monoclinic, $\delta_M$ (red), 
lattice distortions varying $\tilde g_{66}$ and $\tilde g_{12}$ 
at $T=10$~K using the spin-fermion 
model with $J_{\rm NNN}$=$J_{\rm NN}$=0. The scale on the right shows that 
the lattice distortions obtained numerically are within the correct order 
of magnitude when compared with experimental data~\cite{bao,li,huang}. The
values for the orthorhombic distortion are plotted with a negative sign for
simplicity to display.}
\label{distNH}
\end{center}
\end{figure}

In Fig.~\ref{pdt110NH} we display the low-temperature phase diagram 
in the plane $\tilde g_{12}-\tilde g_{66}$ for the case $J_{\rm NN}$=$J_{\rm NNN}$=0. 
Again the collinear and bicollinear phases are stabilized but, 
as expected, smaller values of the monoclinic coupling are needed 
to induce the monoclinic phase. Note, however, that a finite value $\tilde g_{12}\approx 0.1$ 
is still required to stabilize the 
bicollinear phase because the tight-binding term in the Hamiltonian 
still favors a collinear magnetic state via FS nesting. 

The strength of the lattice distortion of Fig.~\ref{pdt110NH} is 
shown in Fig.~\ref{distNH}. A reasonable coupling $\tilde g_{66}\approx 0.2$ 
is needed to reproduce the experimental value of the orthorhombic 
distortion corresponding to the 122 parent compounds. 
The scale shows that the range in the values of the 
stabilized monoclinic distortion is also in 
qualitative agreement with experiments~\cite{bao,li,huang}.

\begin{figure}[thbp]
\begin{center}
\includegraphics[width=8cm,clip,angle=0]{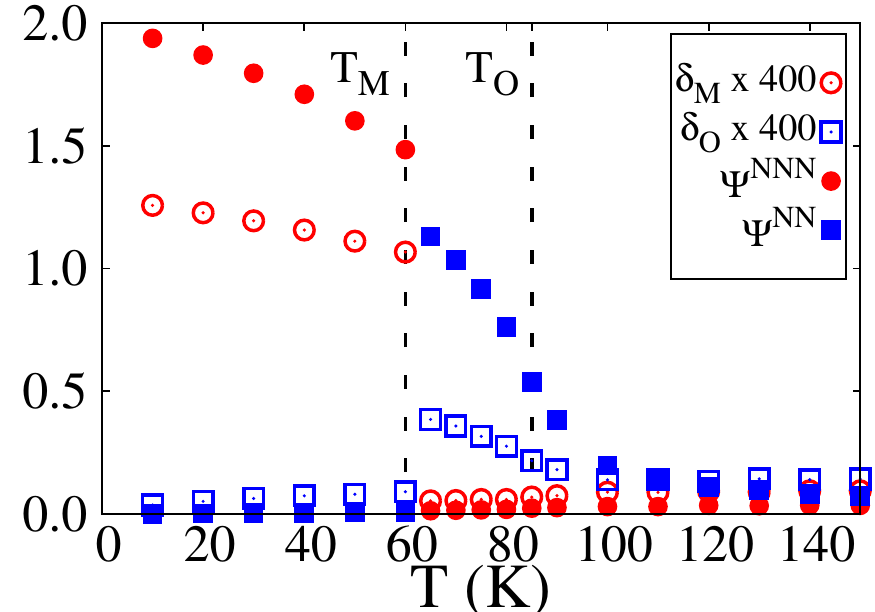}
\vskip -0.3cm
\caption{(color online) Orthorhombic, $\delta_O$ (blue open squares), 
and monoclinic, $\delta_M$ (red open circles), lattice distortions 
and the spin nematic order parameters $\Psi^{NN}$ 
(blue filled squares) and $\Psi^{NNN}$ (red filled circles) as a function of temperature
corresponding to the case 
$\tilde g_{12}=0.29$, $\tilde g_{66}=0.05$ and with the inclusion of Heisenberg couplings.}
\label{opam}
\end{center}
\end{figure}

\section {UNEXPECTED INTERMEDIATE TEMPERATURE RANGE}

When Heisenberg couplings are included, 
the inset of Fig.~2 (main text) shows 
an exotic region where the bicollinear/monoclinic transition 
is preceded by an orthorhombic transition upon cooling.
In Fig.~\ref{opam} we show the magnetic and structural order parameters 
for both types of transitions in this unexpected regime. The transition to the 
collinear/orthorhombic region occurs at about $T=80$~K and
it appears to be continuous, 
while the bicollinear/monoclinic transition occurs at $T=60$~K 
and is strongly first order. Note that in our simulations the 
orthorhombic phase appears to be 
accompanied by a collinear magnetic state while experimentally the orthorhombic 
phase that precedes the monoclinic state in FeTe with excess Fe is magnetically 
incommensurate~\cite{mizuguchi,rodriguez}. We may need either larger lattices 
or the explicit addition of extra irons in order to capture the magnetic 
incommensurability of this phase.

\begin{figure}[thbp]
\begin{center}
\includegraphics[width=8cm,clip,angle=0]{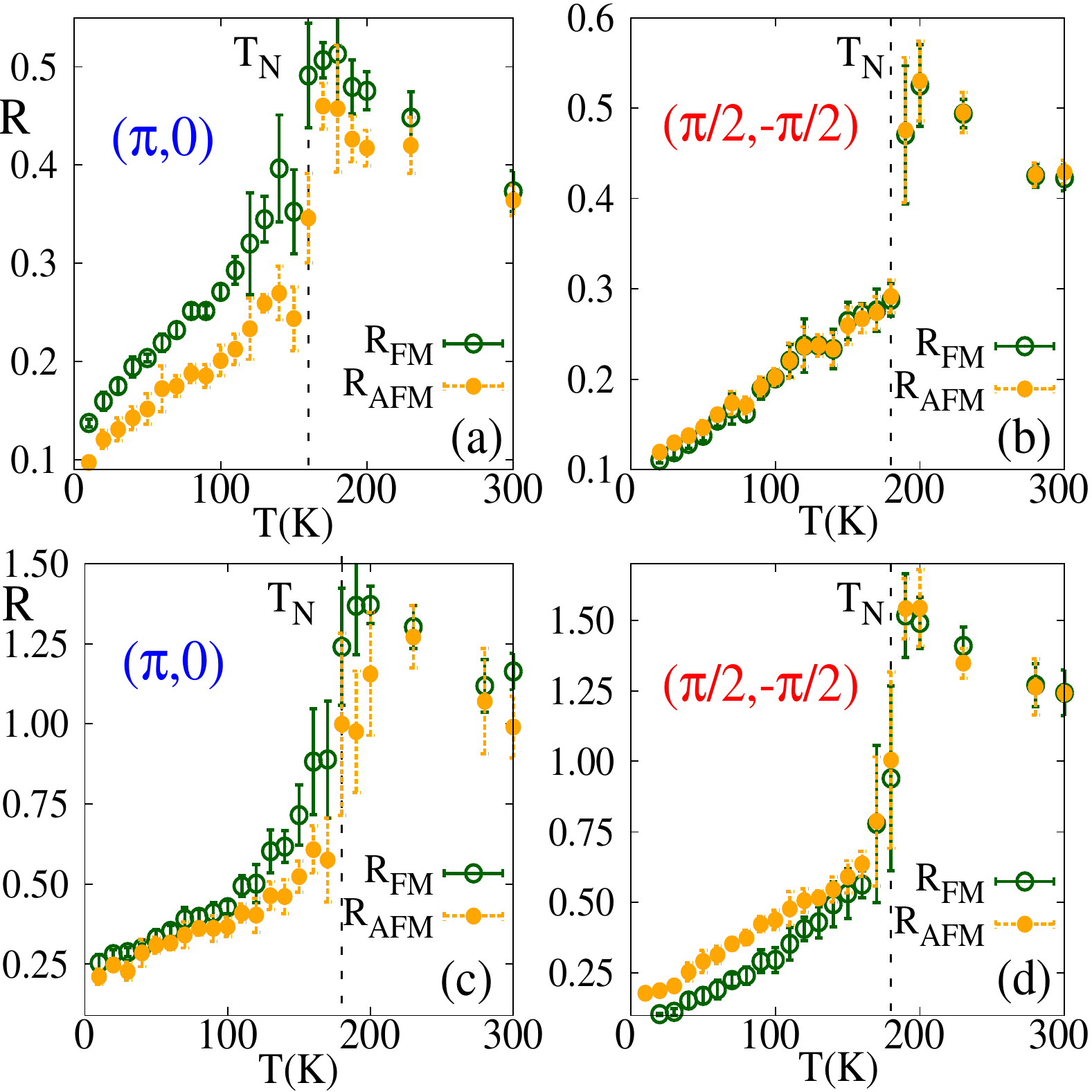}
\vskip -0.3cm
\caption{(color online) Resistance ($h/2e^2$ units) 
vs. temperature along the AFM (orange points) and FM (green points) 
directions in: (a) the collinear/orthorhombic state 
at $\tilde g_{66}=0.16$, $\tilde g_{12}=0.00$, $J_{\rm H}=0.10$~eV, 
and nonzero Heisenberg couplings;
(b) same as (a) but for the bicollinear/monoclinic state with $\tilde g_{66}=0$ and $\tilde g_{12}=0.40$; 
(c) same as (a) but for $J_{\rm H}=0.20$~eV; (d) same as (b) but for $J_{\rm H}=0.20$~eV.} 
\label{RFEASTE}
\end{center}
\end{figure}

\section {REVERSED RESISTIVITY}

A very interesting result that is reproduced by our study
is the anisotropy observed in the planar resistivity of FeTe.

In general, one of the most puzzling behaviors observed 
in the Fe-based materials is the anisotropic 
behavior of the in-plane resistivity as the temperature decreases. In the pnictides the 
cause of the anisotropy is usually attributed to nematicity of electronic origin.
In isovalent or electron 
doped pnictides the resistivity anisotropy develops in the orthorhombic phase 
and the resistivity is lower along the direction with the 
largest lattice constant which becomes 
the antiferromagnetic direction below the magnetic critical temperature. 
This behavior is in principle counterintuitive because in the colossal 
magnetoresistive manganites it is well-known that electrons move 
better in ferromagnetic states. In principle 
this is not the case in the pnictides 
due to the geometry of the orbitals that appear at the Fermi 
surface. 
Interestingly, a ``reversed'' or ``negative''
anisotropy in the resistivity has been observed in the chalcogenides, 
both in the parent compound FeTe~\cite{uchida,jiang} and also in FeSe~\cite{tanatar}.

\begin{figure}[thbp]
\begin{center}
\includegraphics[width=8cm,clip,angle=0]{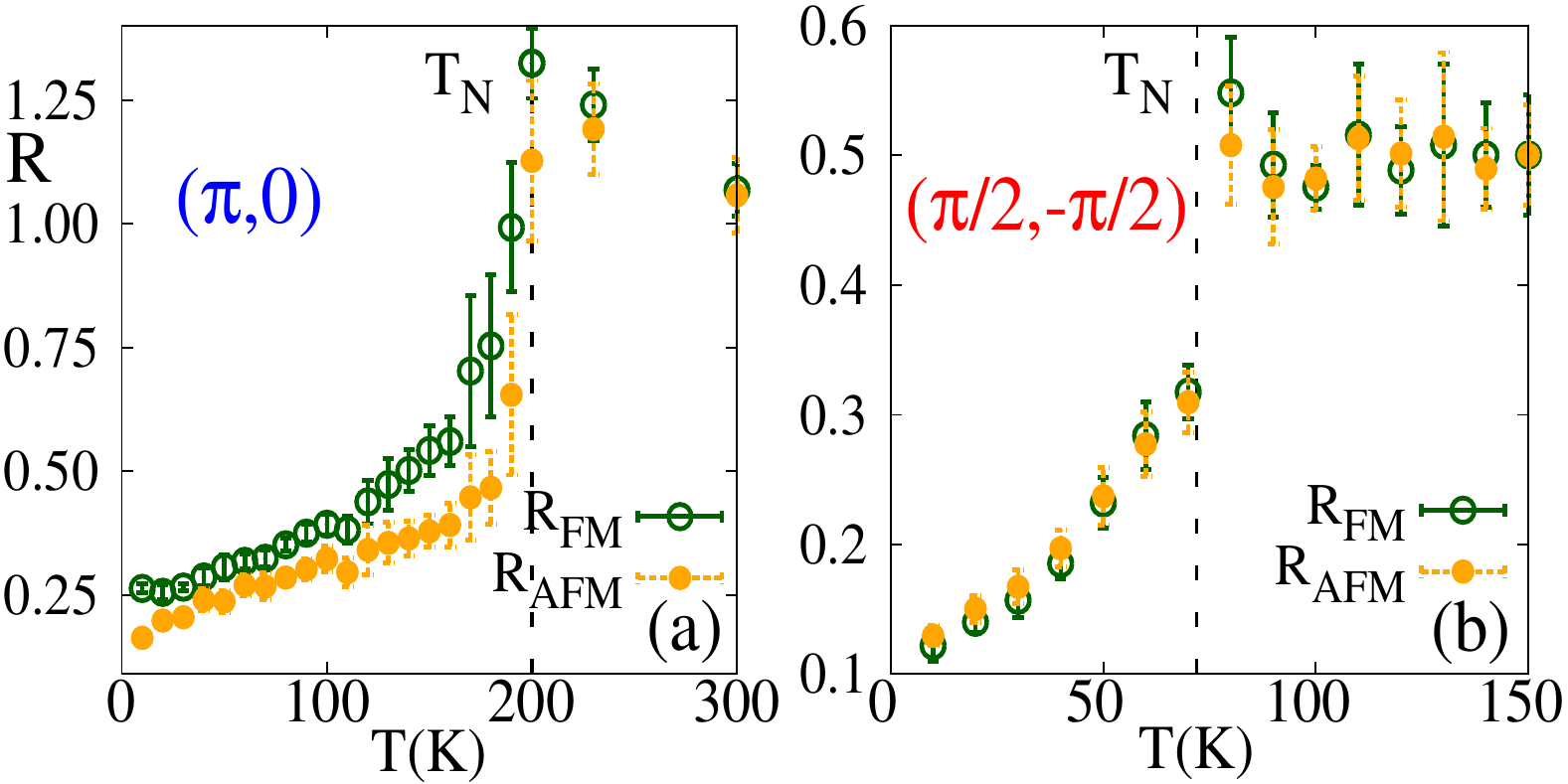}
\vskip -0.3cm
\caption{(color online) Resistance vs. temperature along the AFM (orange points) 
and FM (green points) directions in: (a) the collinear/orthorhombic state 
for $\tilde g_{66}=0.24$, $\tilde g_{12}=0.00$, $J_{\rm H}=0.20$~eV, 
and $J_{\rm NNN}$=$J_{\rm NN}$=0;
(b) same as (a) but for the bicollinear/monoclinic state with 
$\tilde g_{66}=0.00$, $\tilde g_{12}=0.24$, $J_{\rm H}=0.10$~eV, 
and $J_{\rm NNN}$=$J_{\rm NN}$=0.} 
\label{RFEASTENH}
\end{center}
\end{figure}

The resistance $R$ along the AFM and FM directions was calculated as a function of 
the temperature following the procedure described in~\cite{shuhua} 
implementing twisted boundary conditions 
so that the number of accessible momenta along the $x$ and $y$ 
directions was as large as $L=256$. In Fig.~\ref{RFEASTE}~(a) 
we show the planar resistance  in the 
collinear/orthorhombic phase corresponding to $\tilde g_{66}=0.16$, $\tilde g_{12}=0.00$,
$J_{\rm H}=0.10$~eV, and nonzero Heisenberg couplings. In this case, 
the resistance is the smallest along the AFM direction 
($x$-direction in the square lattice) in agreement with previous 
theoretical investigations~\cite{shuhua13} and with the experimental 
data for pnictides~\cite{fisher.science1}. 
In the bicollinear phase, obtained for example at  $\tilde g_{66}=0$ and $\tilde g_{12}=0.40$ 
we actually observe the reversed behavior as shown in Fig.~\ref{RFEASTE}~(b) although here 
the anisotropy is very small~\cite{aniso}. However, it is experimentally known that 
the magnetic moment measured in the chalcogenides 
is larger than the one in the pnictides~\cite{bao,li} and, for this reason, we have repeated the
simulation increasing the Hund coupling from 0.10~eV to 0.20~eV. 
As it can be observed in Fig.~\ref{RFEASTE}~(d) the reversed anisotropy effect is now enhanced.
On the other hand, a similar increase in Hund coupling decreases the resistance anisotropy 
in the orthorhombic phase as shown in panel (c) of the same figure. These results indicate that
the reversed anisotropy is favored (hindered) by the increase (decrease) in the magnitude
of the magnetic moments. A similar response to the Hund coupling is observed for the 
case where the Heisenberg couplings are zero, as presented in the main text: 
in Fig. ~\ref{RFEASTENH} we display the results illustrating 
how the anisotropy is reduced with increasing Hund coupling in the collinear phase (panel a)
while the reversed anisotropy decreases when the Hund coupling is reduced in the bicollinear phase (panel b).

As already explained in the main text, we believe that this ``reversed'' anisotropy
occurs for reasons similar to those unveiled in manganite investigations~\cite{CMR}, namely when
electrons move along the AFM direction they must pay an energy as large as $J_{\rm H}$, while
along the FM direction there is no such penalization. This is compatible with the observation
that the magnitude of the reversed effect increases with $J_{\rm H}$.



\begin{thebibliography}{}
\bibitem{foot} The magnitude of the localized spins is set to $S_{\bf i}=1$ 
since the actual value can be absorbed into the Hamiltonian parameters.
\bibitem{Hubbard} Note that a complete study of the 
bicollinear magnetic state would require a five-orbital Hubbard model
with repulsion $U$ and Hund coupling $J_{\rm H}$, plus the lattice,
all at finite temperature. Such a formidable many-body challenge is not practical,
thus the SF model provides a simplification that
allows for the study of structural, orbital, and magnetic effects simultaneously.
\bibitem{only} The spin in $H_{\rm SLO}$ and $H_{\rm SLM}$ is only 
the localized spin for computational simplicity.
\bibitem{moment} There are two degenerate bicollinear states with momentum $(\pi/2,\pi/2)$ and $(\pi/2,-\pi/2)$ whose degeneracy is broken by the $\mathcal{M}_{\rm ono}$ distortion.
In this work we present results for the case in which the state with momentum $(\pi/2,-\pi/2)$ is stabilized.
\bibitem{comment-10K} The low-temperature (10~K) 
phase diagrams varying $\tilde g_{66}$ 
and $\tilde g_{12}$ are also in the Suppl.
Sec.~\cite{SuSe} [see Fig.~S2 (S3) with (without) Heisenberg couplings]: 
in a broad range of couplings the bicollinear/$\mathcal{M}_{\rm ono}$ 
state is indeed spontaneously stabilized.
\bibitem{hund-aniso} Chalcogenides have higher magnetic moments 
than pnictides, thus likely larger Hund couplings. We also noticed that
for $J_{\rm H}=0.1$~eV we found that the reversed anisotropy was still there 
but smaller (see Suppl. Sec.~\cite{SuSe}).
\end{thebibliography}
\begin{thebibliography}{99}
\bibitem{aniso} Notice that in the bicollinear phase illustrated in Fig.~1 of the main
text, the AFM direction is $x-y$.
\end{thebibliography}
\end{document}